\documentclass[twocolumn]{article}
\usepackage{pgfplots}
\pgfplotsset{compat=newest}
\usepackage[utf8]{inputenc}
\usepackage{microtype}
\usepackage[numbers,square]{natbib}
\usepackage{authblk}

\usepackage{booktabs}
\usepackage{color}
\usepackage{url}
\usepackage{graphicx}
\usepackage{amsmath}

\usepackage{acronym,url}

\usepackage{makecell}
\usepackage{enumitem}
\usepackage{subcaption}

\usepackage{float}

\usepackage{soul}

\acrodef{SASSEC}[SASSEC]{Stereo Audio Source Separation Evaluation Challenge}
\acrodef{BAQ}[BAQ]{Basic Audio Quality}
\acrodef{SiSEC}[SiSEC]{Signal Separation Evaluation Campaign}
\acrodef{SEBASS}[SEBASS]{Subjective Evaluation of Blind Audio Source Separation}
\acrodef{DNN}[DNN]{Deep Neural Network}
\acrodef{PSM}[PSM]{Phase-Sensitive Mask}
\acrodef{STFT}[STFT]{Short-Time Fourier Transform}
\acrodef{DS}[DS]{Dialogue Separation}
\acrodef{DE}[DE]{Dialogue Enhancement}
\acrodef{SD}[SD]{Standard Deviation}

\newcommand\numExpertsOrig{26}

\newcommand\numExperts{42} 
\newcommand\numScores{10080}

\title{Expanding and Analyzing ODAQ -- the Open Dataset of Audio Quality}

\author[1]{Sascha Dick}
\author[2]{Christoph Thompson}
\author[3]{Chih-Wei Wu}
\author[1]{Matteo Torcoli}
\author[1]{Pablo Delgado}
\author[3]{Phillip A. Williams}
\author[1]{Emanu\"{e}l Habets}

\affil[1]{Fraunhofer Institute for Integrated Circuits IIS, Erlangen, Germany}
\affil[2]{Ball State University, Muncie, USA}
\affil[3]{Netflix, Inc., Los Gatos, USA}

\usepackage[pscoord]{eso-pic}
\newcommand{\placetextbox}[3]{% \placetextbox{<horizontal pos>}{<vertical pos>}{<stuff>}
\setbox0=\hbox{#3}% Put <stuff> in a box
\AddToShipoutPictureFG*{% Add <stuff> to current page foreground
\put(\LenToUnit{#1\paperwidth},\LenToUnit{#2\paperheight}){\vtop{{\null}\makebox[0pt][c]{#3}}}%
}%
}%

\date{}

\begin{document}

\twocolumn[
\maketitle
\begin{abstract}
The Open Dataset of Audio Quality (ODAQ) was recently introduced to address the scarcity of openly available audio datasets with corresponding subjective quality scores.
The dataset, released under permissive licenses, comprises audio material processed using six different signal processing methods operating at five quality levels, along with corresponding subjective test results. 
To expand the dataset, we provided listener training to university students to conduct further subjective tests and obtained results consistent with previous expert listeners. We also showed how different training approaches affect the use of absolute scales and anchors.
The expanded dataset now comprises results from three international laboratories providing a total of \numExperts\ listeners and \numScores\ subjective scores. This paper provides the details of the expansion and an in-depth analysis. As part of this analysis, we initiate the use of ODAQ as a benchmark to evaluate objective audio quality metrics in their ability to predict subjective scores.
\end{abstract}
\vspace{12pt}
]

\placetextbox{0.5}{0.08}{\fbox{\parbox{\dimexpr\textwidth-2\fboxsep-2\fboxrule\relax}{\footnotesize \centering Accepted for presentation at the Audio Engineering Society (AES) 157th Convention, October 2024, New York, USA.}}}

\section{Introduction}

Datasets of processed audio signals and subjective quality scores are instrumental for perceptual research and reproducible evaluation of objective audio quality metrics. However, openly available datasets are scarce due to listening test efforts and copyright concerns limiting the distribution of audio material in existing datasets.
For instance, in \cite{dick2017}, we generated various types of isolated signal degradations and typical audio coding artifacts. We evaluated their impact on the perceived quality using subjective listening tests and published the test results. In this case, the employed audio signals could not be publicly shared due to copyright limitations. Another example is the USAC verification test~\cite{n12232} dataset, which has been widely used (e.g., \cite{Torcoli2021, biswas2023, DelgadoPEAQ})
but is also not publicly available.

To address this issue and create a common ground for the research community, we have introduced the Open Dataset of Audio Quality (ODAQ)\footnote{\url{https://github.com/Fraunhofer-IIS/ODAQ/}\label{first_footnote}}. In its first version\footnote{\url{https://doi.org/10.5281/zenodo.10405774}}~\cite{Torcoli2024ODAQ}, ODAQ contains subjective test results from \numExpertsOrig\ expert listeners from two international laboratories, along with audio material under permissive licenses. The employed user interface for conducting subjective listening tests was also made available as an open-source tool\footnote{\url{https://github.com/Netflix-Skunkworks/listening-test-app}\vspace{0.25em}}. 
Various audio signals were collected with audio for broadcast and streaming in mind while also taking great care to select only material with permissive licenses. The selected signals were then processed by audio processing methods representing audio coding as in \cite{dick2017}, as well as source separation for \ac{DE}.

With ODAQ, we established a foundational dataset that can be employed for verification and benchmarking of objective quality metrics while also serving as a common ground for psychoacoustic research. However, this is only an initial step in addressing the shortage of openly-available subjective data. In this contribution, we expand the dataset with additional subjective scores and demonstrate how consistent subjective results can be obtained by training listeners to be expert listeners.
We also present additional details on the signal creation and discuss adjustments that have been made compared to our previous tests in \cite{dick2017}.
Furthermore, we investigate the relation between the parameters used to generate the signal degradations and the corresponding subjective scores, as well as the differences between listener groups.
Finally, we showcase how ODAQ could evolve as a shared resource for the research community by integrating new contributions, and we leverage the expanded dataset for benchmarking existing objective metrics. In total, the expanded ODAQ now includes \numScores\ subjective scores from \numExperts\ expert listeners assessing \ac{BAQ} scores via MUSHRA~\cite{MUSHRA}. The additional subjective scores are available online\footnote{\url{https://doi.org/10.5281/zenodo.13377284}\vspace{0.25em}}, and the example code for benchmarking objective metrics are available in the ODAQ online repository$^{\ref{first_footnote}}$.
\section{Test Material}
\label{sec:testmaterial}
For the expansion of ODAQ, we used the same set of test materials as described in \cite{Torcoli2024ODAQ}. 
The following section provides additional details and discusses the signal selection and generation of artifacts.

\subsection{Choice of Audio Material}
The audio material was selected to cover a wide range of signal properties, including music and broadcast-like content and critical content that is known to provoke specific types of signal degradation or coding artifacts.
The content consists of 11 movie-like soundtracks comprising dialogues mixed with music and effects and 14 music excerpts, out of which 8 are solo instrument recordings.

As discussed in \cite{dick2017}, specific "pure" coding artifacts are most prominent for specific signal types (e.g., tonal solo instruments for tonal artifacts or transient signals for temporal artifacts), whereas "mixed" content will typically elicit a combination of different artifact types. Therefore, in \cite{dick2017}, different signals were selected for each artifact type to maximize control over the artifact generation. One limitation of this approach is the inability to compare the ratings across different artifact types directly. As a compromise, in ODAQ, we selected two "pure" signals per artifact type. In addition, two "mixed" music signals were processed with all coding artifact types. A detailed list with a description of the test signals is given in Table~\ref{tbl:items}.

\subsection{Audio Coding Artifacs}
\label{sec:aca}

The test signals in ODAQ \cite{Torcoli2024ODAQ} were generated following the method presented in \cite{dick2017} to generate the five isolated types of  audio coding artifacts by forcing audio coders into controlled, sub-optimal operation modes:%

\begin{enumerate}[itemsep=0pt,topsep=0pt, leftmargin=*]
\item \textbf{Low-Pass (LP)}: Bandwidth Limitation (BL) in~\cite{dick2017}.
\item \textbf{Tonality Mismatch (TM)}: noise-like components are substituted with tonal components.
\item \textbf{Unmasked Noise (UN)}: tonal components are substituted by noise with the same spectral envelope. 
\item \textbf{Spectral Holes (SH)}: spectral parts are quantized to zero (here with fixed, controlled probability).
\item \textbf{Pre-Echoes (PE)}: temporal smearing of quantization noise around transients.
\end{enumerate}

All items were normalized to $-23$\,LUFS integrated loudness~\cite{ebu_r128} before processing. 

Table~\ref{tbl:artifacts} provides an overview of the processing methods and the parameters used to generate the 5 different quality levels.

The processing methods simulate artifacts that may occur in suboptimal waveform-preserving audio coders (LP, SH, and PE), as well as artifacts possibly due to parametric coding approaches, such as bandwidth extension at very low bitrates (TM, UN) with low crossover frequencies.

\begin{table}[!t]
\begin{footnotesize}
\resizebox{\columnwidth}{!}{%
\centering
\setlength{\tabcolsep}{4pt}

\begin{tabular}{l l}
\toprule
\textbf{item} & \textbf{description}  \\ \midrule
%\multicolumn{2}{|c|}{{mixed} music items} \\ \hline
MIX1  & Upbeat pop music \\ %\hline
MIX2  & Percussive music with solo flute \\ %\hline 
%\hline \multicolumn{2}{|c|}{{pure} TM items} \\ \hline
%\hline
TM1 & Solo trumpet playing medium tempo accentuated notes\\ %\hline
TM2 & Solo violin playing slow, held notes  \\ %\hline
%\hline \multicolumn{2}{|c|}{{pure} SH items} \\ \hline
%\hline
SH1 & Mixed choir singing choral music \\ %\hline
SH2 & Solo glockenspiel playing distinct notes \\ %\hline
%\hline \multicolumn{2}{|c|}{{pure} LP items} \\ \hline
%\hline
LP1 & Finger picking on solo acoustic guitar \\ %\hline
LP2 & Jazz trumpet solo over band accompaniment \\ %\hline
%\hline \multicolumn{2}{|c|}{{pure} UN items} \\ \hline
%\hline
UN1 & Solo accordion playing melody over rhythmic chords \\ %\hline
UN2 & Solo violin playing mid-tempo melody \\ %\hline
%\hline \multicolumn{2}{|c|}{{pure} PE items} \\ \hline
%\hline
PE1 & Solo castanets playing slow and fast\\ %\hline
PE2 & Hand clapping by single person, distinct claps \\ %\hline
%\hline \multicolumn{2}{|c|}{{pure} DE items} \\ \hline
%\hline
DE1 & Male speech over seaside noises \\ %\hline
DE2 & Male speech over ambient music \\ %\hline
DE3 & Male speech over metallic effects and ambient music \\ %\hline
DE4 & Male speech over suspenseful music \\ %\hline
DE5 & Male speech over smooth jazz\\ %\hline
DE6 & Male speech over applause \\ %\hline
DE7 & Female and male speech over crowd noise \\ %\hline
DE8 & Female speech over percussive music \\ %\hline
DE9 & Female speech over cinematic music \\ %\hline
DE10 & Female speech over string music\\ %\hline
\bottomrule
\end{tabular}
}
\captionof{table}{\label{tbl:items} List of test stimuli in ODAQ.}
\end{footnotesize}
\end{table}
\begin{table}[!t]

%\begin{table}[b]
\begin{footnotesize}
\resizebox{\columnwidth}{!}{%
\centering
\setlength{\tabcolsep}{4pt}
\begin{tabular}{c c c c c c c}
\toprule
\textbf{Method} & \textbf{Parameter} & \multicolumn{5}{c}{\textbf{Quality Level}} \\[2pt] \cmidrule{3-7}
\textbf{} & \textbf{} & \textbf{Q1} & \textbf{Q2} & \textbf{Q3} & \textbf{Q4} & \textbf{Q5}\\ \midrule
LP & freq. [kHz]  & $5.0$ & $9.0$ & $10.5$ & $12.0$ & $15$ \\ 
TM & freq. [kHz] & $3.0$ & $5.0$ & $7.0$ & $9$ & $10.5$ \\ 
UN & freq. [kHz] & $3.0$ & $5.0$ & $7.0$ & $9.0$ & $10.5$ \\ 
SH & hole prob. [\%] & $70$ & $50$ & $30$ & $20$ & $10$ \\ 
PE & \makecell{NMR[dB]\\length}& \makecell{10\\4096}  & \makecell{10\\2048} &  \makecell{10\\1024} & \makecell{16\\2048} & \makecell{16\\1024} \\ 
DE & DS system & \cite{openunmix} & \cite{TFCTDF} & \cite{Petermann22} & \cite{DeepFilterNet2} & q-PSM \\ \bottomrule
\end{tabular}
}
\captionof{table}{\label{tbl:artifacts}Processing methods and quality levels used to generate the conditions under test.}
\label{TableArtifactParameters}
\end{footnotesize}
%\end{table}
\end{table}

The subjective results in \cite{dick2017} showed good coverage of the quality scale but exhibited non-uniform distribution in the lower quality levels. Therefore, the parametrization of the artifact generation for ODAQ was adjusted based on the findings in \cite{dick2017} and on preliminary listening sessions carried out by the authors of \cite{Torcoli2024ODAQ}:
\begin{itemize}[itemsep=0pt,topsep=0pt, leftmargin=*]
    \item For frequency-based artifacts (LP, TM, and UN), the cross-over frequencies were adapted to be more evenly spaced in the lower frequency range and coarser spaced in the higher frequency range. 
    As LP is generated the same way as the low-pass anchors in MUSHRA, a finer spacing with two additional levels (3.5 and 7\,kHz) is already present for this artifact.

    \item For PE, two different aspects can impact perception: The \emph{amount} of temporally smeared quantization noise, as well as the \emph{temporal extent} of the smearing. For finer control over the influences on quality levels, the Noise-to-Mask Ratio (NMR) \cite{brandenburg1987evaluation} as well the transform block length were both varied to obtain independent control over both aspects.
    
    \item Although the focus of the generated artifacts in~\cite{dick2017} was on monaural signal degradations, listeners reported cases where the stereo image was also affected. This occurred especially for SH due to the channel-independent generation of spectral holes and, to a lesser extent, for PE due to the spatial widening of the introduced noise. To mitigate the introduction of spatial artifacts, we employed Mid-Side (MS)-Stereo processing before the artifact generation for SH and PE. The Mid and Side signals were both individually processed by the respective methods and then transformed back into left and right signals, resulting in artifacts perceived as coming from the phantom center of the stereo image. 
\end{itemize}

\subsection{Dialogue Enhancement (DE)}
\label{sec:ds}

\ac{DE} is a solution that allows the audience of TV and streaming services to personalize the audio mix and adapt the relative level of speech (or \textit{dialogue}) to their needs and preferences \cite{fuchs2012dialogue,ward2019casualty,torcoli2021dialog,amazon2023}. When the clean speech signal is not separately available, source separation techniques (often referred to as \ac{DS}) can be employed to estimate the speech signal from the full audio track and so enable \ac{DE}. 

In ODAQ, we considered four separation methods based on \acp{DNN}, for which implementations and pre-trained networks are publicly available: 
Open Unmix \cite{openunmix},
TFC-TDF Unet \cite{TFCTDF},
Cocktail Fork Baseline Model \cite{Petermann22}, 
\mbox{DeepFilterNet 2}~\cite{DeepFilterNet2}, and an oracle-knowledge approach using
a quantized Phase-Sensitive Mask (q-PSM), as described in \cite{Torcoli2024ODAQ}. Preliminary listening by the authors led to the selection of these separation systems from a longer list of candidate systems. The criterion for selection was to obtain 5 distinct quality levels spanning a significant portion of the MUSHRA scale. This was successfully achieved, as later confirmed by the listening tests.

The estimated dialogue signals are re-mixed with the corresponding background estimates, attenuating the background by $20$~dB. The obtained stereo mixtures were normalized to $-23$~LUFS integrated loudness.

\section{Listening Tests}

\subsection{Test Method and Setup}
To ensure consistency with the previous experiments, the additional results presented in this paper were obtained using the same ODAQ v1 package, the same procedure, and the same software outlined in \cite{Torcoli2024ODAQ}.
The procedure considers the five different quality levels of a processing method within each MUSHRA test trial. Along with them, a hidden reference and 3.5\,kHz and 7\,kHz low-pass anchors are presented, resulting in 8 conditions to grade per trial, and a total of 30 trials.
As also done in \cite{Torcoli2024ODAQ}, the tests were divided into 3 sessions with 10 trials each to avoid listener fatigue. Furthermore, the listeners were advised to take sufficient breaks between the sessions. The order of the trials was randomized across all sessions for each listener.

Listeners were provided with written instructions detailing the test procedures.
Before the actual test, each listener participated in a brief training session to familiarize themselves with the methodology, user interface, and test conditions. The training session includes three trials with audio samples that are not used in the actual test. Listeners were instructed to adjust the playback volume to a comfortable level during training and avoid further adjustments in the actual test.

All listening tests were conducted in acoustically damped listening rooms that are suitable for audio mixing or screening. For playback, Beyerdynamic DT770 Pro 250 Ohm closed-back headphones were used in combination with a professional 24-bit soundcard.

\subsection{Test Subjects and Listener Training}

\subsubsection{Test Subjects}

The initial pool of listeners in ODAQ as presented in \cite{Torcoli2024ODAQ} consisted of expert listeners and experienced audio coding experts in two laboratories in Germany and the United States (US). Here, this combined pool of listeners will be referred to as \emph{Cohort A}.

To expand the pool of listeners, expert listener training was followed by a listening test conducted at a third laboratory at Ball State University (BSU) in the US. 
The test subjects were undergraduate students enrolled in either the Telecommunications program in the College of Media or the Music Media Production program in the School of Music. The Music Media Production students are trained musicians and receive musical ear training in addition to the technology courses in the curriculum. Neither group of students was experienced in taking MUSHRA tests. To investigate the impact of listener training, two different approaches were employed to train the participants, resulting in listener groups that will be referred to as \emph{Cohorts B1} and \emph{B2}.

Cohort A had a combined 26 subjects with a mean age of 37.5 years (Standard Deviation (SD): 8.5) with an average of  12.7 (SD: 10) years of professional audio experience in areas such as audio mixing/production, music performance, or audio research/development.
Cohort B1 had eight subjects with a mean age of 20.4 years (SD 0.34 years), with an average of 2.5 years of experience (SD 0.76). Cohort B2 had eight subjects with a mean age of 20.75 years (SD 2.22 years), with an average of 2 years of experience (SD 0.97).

\subsubsection{Expert Listener Training}
\label{sec:listenertraining}
Listener Cohorts B1 and B2 were trained using encoded stimuli ranging in quality from transparent to medium-quality audio coding and a training set of stimuli with deliberately generated artifacts. Good training results can be achieved by exposing listeners to highly specific impairments generated directly for training purposes, e.g., as presented in \cite{herre2023}.
Group training sessions were conducted in a treated studio space and a stereo pair of Genelec 1037 speakers. The group size was limited to 4 students.

The ITU-R recommendation BS.1534-3 on conducting MUSHRA tests \cite{MUSHRA} states that the current standardized anchors for the MUSHRA tests may not be appropriate for all systems under test. 
To investigate the potential impact of anchor usage, the administrator for Cohorts B1 and B2 introduced a small variation to modify the training outlined in BS.1534-3 between Cohorts B1 and B2, which otherwise both consisted of listeners all from the same experience level, age range, and educational background.
The training for Cohort B2 did not address the concept of the anchor, and no use of the term was made during oral instruction. Both cohorts were exposed to the same training materials and levels of impairment, including the 3.5\,kHz, and 7\,kHz anchor conditions, but only for Cohort B1 the anchor condition was referred to as such and introduced under the explicit anchor label when played for the group.

Both cohorts received three group training sessions over the course of two weeks and one final training session individually with the testing app. 
In Session 1, participants were introduced to general coding artifacts, limited to easy and obvious impairments. Encodes were generated with state-of-the-art codecs at low-quality bitrates. In Session 2, uncompressed audio was compared to mid- and high-quality bitrates, and an informal assessment through the instructor was done. The third group session focused on specific impairments that would be used in the test. After an informal assessment through the instructor, open discussion between participants was encouraged to explore listening strategies. The session was concluded after an introduction to the MUSHRA application and demonstration.
The final training session was congruent with the instructions for the familiarization phase provided in Attachment 1 to Annex 1 in   \cite{MUSHRA}, except for the omission of the term anchor for Cohort B2.

\begin{figure}[htb!]
\centering
\centerline{\includegraphics[width=\linewidth]{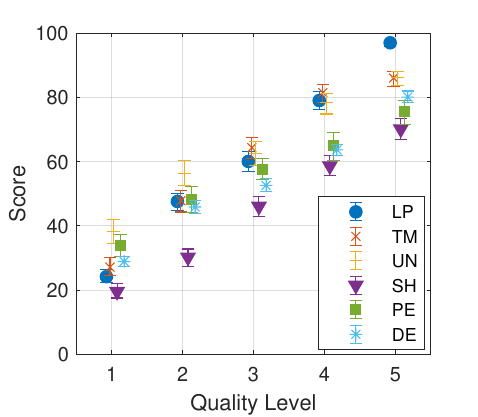}}
\caption{Overall results (\numExperts\ participants) showing mean scores and 95\% confidence intervals (CI) per processing method and quality level.}
\label{fig:results:overall}
\end{figure}

\section{Results and Discussion}

\subsection{Overall Quality Levels and Scale Coverage}

Across all cohorts, a total of \numExperts \ listeners after post-screening with a total of \numScores\ scores were collected. 
Figure~\ref{fig:results:overall} shows the mean scores and 95\% confidence intervals for the overall results for the different quality levels and processing methods. 
The results show that, overall, a good coverage of the quality scale was achieved for all processing methods. Via the adjustments in parametrization compared to the previous tests in \cite{dick2017}, a more uniform distribution across the quality levels was also successfully achieved. In contrast, the test in \cite{dick2017} showed saturation effects for the PE, UN, and TM artifacts from around quality level~3.

\begin{figure*}[h]
\centering
\centerline{\includegraphics[width=0.92\linewidth]{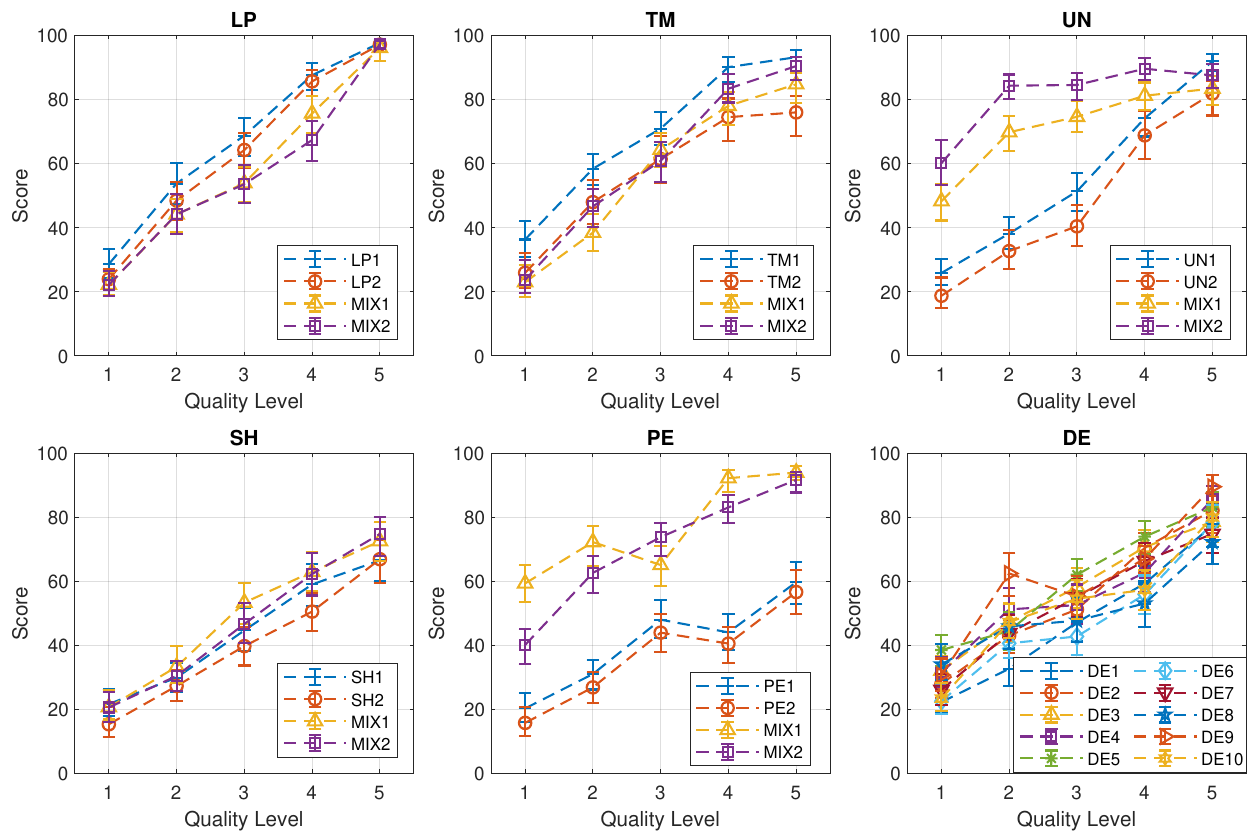}}
\caption{Overall results (\numExperts\ participants) showing mean scores and 95\% confidence intervals (CI) per processing method and quality level for individual test stimuli (see Tables \ref{tbl:artifacts}, \ref{tbl:items}). For PE and UN, the scores of the dedicated critical signals are significantly lower than for the generic, mixed signals.}
\label{fig:results:itemwise}
\end{figure*}

\subsection{Artifact and Item Dependency}
Figure~\ref{fig:results:itemwise} shows the overall results for the different processing methods and quality levels for the individual stimuli (Table~\ref{tbl:items}). 
For the majority of the processing methods, i.e.,\ DE, LP, SH, and TM, there is relatively little dependency on the signal under test. However, for PE and UN, there are significant differences between the \emph{mixed} and the \emph{pure} items across most quality levels. For PE, the overall rating of the mixed items exhibits an offset by ca.\ 20 MUSHRA points higher than for the pure items along the entire quality scale. This indicates that the presence of additional background signals around transients substantially contributes to masking the temporally smeared noise and thus mitigating the perceptibility of PE.
For UN, the quality scores of the mixed items are also overall higher than for the pure items, and they saturate towards the maximum achieved quality already around quality level~2. This is likely caused by the lack of distinct high-frequency tonal components, which reduces the impact of signal components being substituted by noise.

These findings show that even for a controlled parametrization, the impact on the quality score of a given processing method greatly depends on the test signal. This illustrates the challenges in encoder tuning and the development of objective quality metrics.

\subsection{Temporal Extent and Amount of PE}
\begin{figure}[t]
\centering
\centerline{\includegraphics[width=0.75\linewidth]{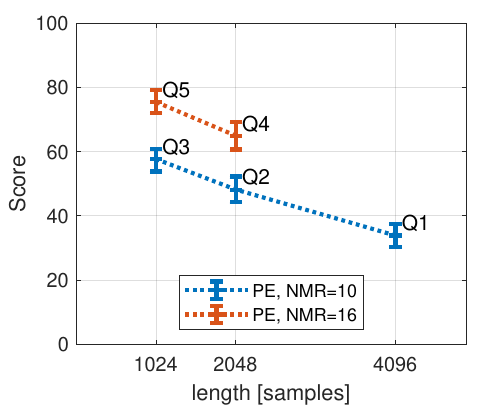}}
\caption{Overall results (\numExperts\ participants) showing mean scores and 95\% confidence intervals (CI) for PE for different NMR and transform length.}
\label{fig:results:PE}
\end{figure}

For PE, two parameters were used to independently control the \emph{amount} (NMR) and the \emph{length} of temporal smearing of quantization noise. Figure~\ref{fig:results:PE} shows the subjective scores, depending on the transform length, for the two different levels of NMR that were used. 
The transform length has a substantial impact on the perceived quality while changing the NMR results in an overall offset of the length-dependent quality score. The longer temporal smearing reduces the amount of temporal masking and is presumably also more likely to be perceived as an additional signal rather than a distortion of the test signal and, thus, more obvious.

\subsection{Listener Groups}
\begin{figure*}[hbt!]
\centering
{\includegraphics[width=\linewidth]{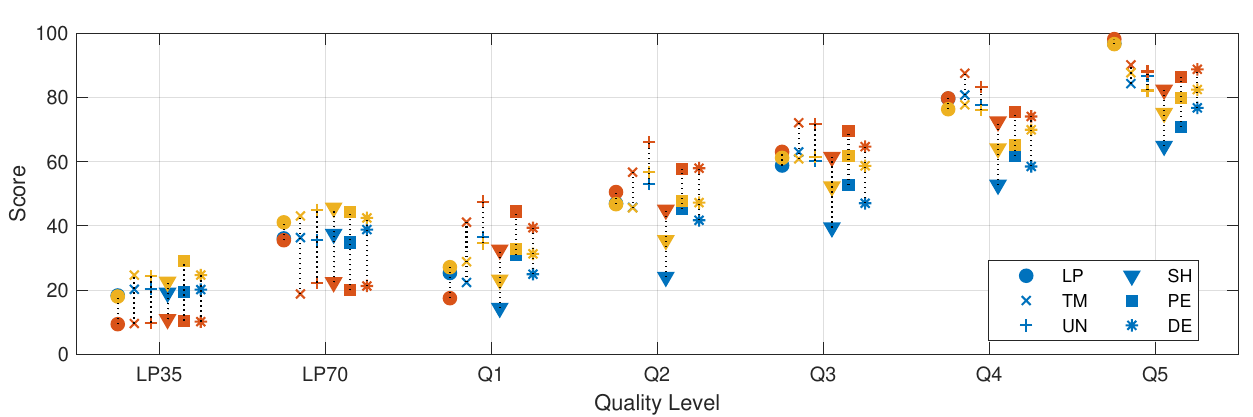}}
\caption{Overall results (\numExperts\ participants) showing mean scores per processing method and quality level for different listener groups (blue: Cohort A, red: Cohort B1, yellow: Cohort B2).}
\label{fig:results:labs}

\centering
\centerline{\includegraphics[width=\linewidth]{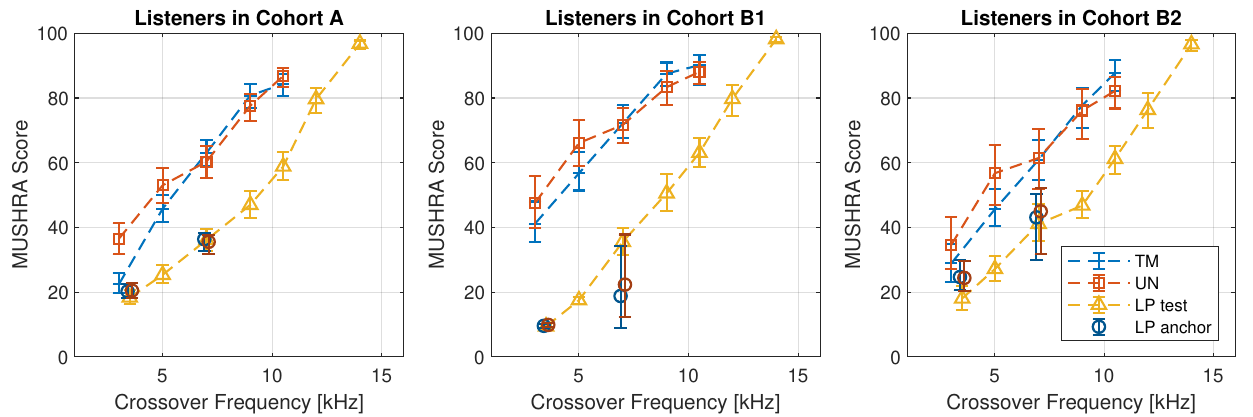}}
\caption{Results (\numExperts\ participants) showing mean scores and 95\% confidence intervals (CI) for frequency-based processing method (TM,UN,LP) depending on crossover frequency, and the rating of the respective low-pass anchors that were presented in the context of the other processing methods. For Cohort B1, low-pass conditions were rated substantially lower when presented as anchor than when presented in the context of other low-pass frequencies, which can be explained by the explicit mention of "anchors" during training.}
\label{fig:results:frequency}
\end{figure*}

To investigate differences between listener groups as well as the potential impact of different listener training approaches (see Section~\ref{sec:listenertraining}), Figure~\ref{fig:results:labs} shows the mean scores of the
different test conditions for each individual listener cohort as indicated by the marker color. 
Overall, the recently trained listeners in Cohorts B1 and B2 show a slight offset towards higher scores compared to Cohort A, but otherwise exhibit the capability to distinguish different quality levels and provide results consistent with the more experienced listeners of Cohort A.

\subsection{Frequency-Dependent Artifacts and Anchors}

For three of the processing methods (TM, UN, and LP), the common control parameter is the crossover frequency. This also allows us to put the frequency-dependent processing conditions in relation to the low-pass anchors that were included in all tests. The subjective quality scores in relation to the crossover frequency are shown in Figure \ref{fig:results:frequency} for the three listener groups of Cohorts A, B1, and B2. For the LP condition, the low-pass anchors are, in principle, just additional steps in the scaling of the control parameter and, therefore, shown in line with the other quality levels. For TM and UN, the ratings of the low-pass anchors in the respective context are shown separately.

The results for Cohort B1 show that 3.5\,kHz and 7\,kHz low-pass was rated over 10 points lower when it served as distinct "anchor" presented in the context of other processing methods than when presented along other low-pass frequencies. For the other listener groups the rating of the low-pass anchor conditions is mostly consistent, regardless of the presentation context.
Cohort B1 showed a bias towards grading the anchors at the bottom of the scale and everything else above, resulting in a shift of overall scale usage. Cohort B2 did not exhibit this bias, and the results were more congruent with those of Cohort A. It is important to point out that the ability to identify impairments in Cohort B1 was not affected by the bias; the reference identification or grading of non-anchor conditions was done with equal consistency. It is also of note that Cohort B1 only graded the 7\,kHz LP anchor higher and consistent with the other LP conditions and more congruent with Cohort A when LP was the condition under test.

\subsection{Interpretation of the Bias in Cohort B1}
Bias in grading behavior in subjective listening tests has long been established. Zielinski \cite{zielinski:2008} found that English-speaking subjects exhibited a clustering of the two lowest ratings of the MUSHRA scale as opposed to German-speaking subjects. His example shows a clustering similar to the Cohort B1 results.

The undue low rating of the 7\,kHz midrange anchor in Cohort B1, which led to the clustering of the 3.5 and 7\,kHz anchors, also invokes the anchoring bias outlined by Kahneman \cite{tversky1974judgment}. This effect is characterized by a subject’s dogmatic adherence to a known reference and utilizing this systematically in the decision-making process.

Considering the strong effect that the modification in training had on Cohort B2  and taking the current questions regarding the suitability of the standard MUSHRA anchors into account, some assumptions can be made: The phrasing in \cite{MUSHRA} may have an influence on anchor grading behavior. The recommendation states: “You should listen to the reference, anchor, and all the test conditions by clicking on the respective buttons.” Any thorough training naturally needs an introduction of the anchor conditions to the listeners. The question becomes whether these need to be labeled as such. The main differences between Cohort A and Cohorts B1/B2 were age and experience in test taking. The median number of MUSHRA tests taken by Cohort A was 10, while that of Cohorts B1 and B2 was 0. This may indicate that listeners who are well trained to identify impairments with high precision but not experienced MUSHRA test participants may exhibit a bias in their anchor grading behavior when the ITU document training procedure and terminology are strictly followed.

\section{Objective Quality Metrics}
\label{sec:objective_quality_metrics}
\subsection{Prediction Performance}

One of the goals for ODAQ is to serve as a shared resource that could be expanded and utilized by the research community and facilitate studies around audio quality. 
Several previous works analyzed the performance of objective quality metrics in predicting ground-truth subjective quality scores, e.g., \cite{Torcoli2021, DelgadoPEAQ, torcoli2018comparing, hu2007evaluation}.
This type of analysis has two issues: 1) It has limited reproducibility since openly available datasets of subjective scores were scarce before ODAQ, and 2) It cannot include newer metrics proposed after the time of writing.
While ODAQ alleviates the first issue, we would also like to address the second one by kicking off an online evaluation\footnote{https://paperswithcode.com/dataset/odaq-open-dataset-of-audio-quality} that can be constantly updated with the newest developments in the field.

As an initial set of benchmark results, we evaluate various systems by comparing their predicted scores against the mean subjective scores in our expanded dataset (i.e., the mean is computed over the scores from all Cohorts A, B1, and B2). 

As a starting point, we consider the following objective metrics:
NMR~\cite{brandenburg1987evaluation} (as implemented within PEAQ \cite{PEAQ, Kabal}), PEAQ ODG \cite{PEAQ, Kabal},
PEAQ-CSM~\cite{delgado2022data},
\mbox{2f-model}~\cite{ksrWaspaa19}, 
ViSQOLAudioV3 \cite{Visqol3}, 
SMAQ \cite{wu2021}, 
wideband PESQ \cite{PESQ, miao_wang_2022_6549559},
SI-SDR \cite{roux2018sdrhalfbakeddone}, and
\mbox{DNSMOS}~\cite{reddy2022dnsmos}. For DNSMOS, we report the performance of the estimated overall quality (OVRL) output only; the signal quality (SIG) output has shown similar performance.
For the metrics supporting only mono inputs, the signals are passively downmixed to mono before measurement.

For each system, the Pearson's correlation coefficients ($R$) \citep{EvalObjective} of all six processing methods (Table \ref{tbl:artifacts}) are first calculated and reported separately.
As in \cite{Torcoli2021}, scores for anchor and reference conditions are not considered in the correlation analysis.
In addition, an aggregated score is reported to characterize the overall performance of each objective metric.
To compute the aggregated score, Fisher's Z-transform is applied to the correlation coefficients of each metric and processing method; the mean value is computed in the transform domain; and the inverse transformation is applied \cite{Torcoli2021}. 

\subsection{Prediction Performance Results}

\begin{figure}[t]
\resizebox{0.49\textwidth}{!}{%
\begin{tikzpicture}
\begin{axis}[
    axis on top,
    %axis line style={draw=none},
    %tick style={draw=none},
    xticklabels={DE, LP, PE, SH, TM, UN, AGG},
    yticklabels={DNSMOS, SI-SDR,PESQ, SMAQ, ViSQOLAudioV3, PEAQ (ODG),2f-model, PEAQ-CSM, NMR},
    xlabel={Processing Method},
    ylabel={Objective Quality Metric},
    xlabel style={yshift=-1em},
    ylabel style={yshift=1em},
    xtick=data,
    ytick=data,
    colormap={mycolormap}{
        color=(red),
        color=(red),
        color=(red!30!white),
        color=(green),
    },
    colorbar,
    point meta min=0, % Set the minimum value of the colormap
    point meta max=1, % Set the maximum value of the colormap
    colorbar style={
        xlabel=$R$,
        xlabel style={yshift=-1em}
    },
    xticklabel style={rotate=90}, % Rotate x labels vertically
    nodes near coords,
    nodes near coords style={
        font=\small,
        font=\bfseries,
        anchor=center,
        /pgf/number format/fixed
    },
    point meta=explicit,
    point meta rel=per plot,
]
\addplot[
    matrix plot*,
    mesh/cols=7,
    point meta=explicit,
    mesh/rows=9
] coordinates {

    (0,1) [0.35] %DE 
    (1,1) [0.19] %LP
    (2,1) [0.65] %PE 
    (3,1) [0.47] %SH 
    (4,1) [0.22] %TM 
    (5,1) [0.33] %UN 
    (6,1) [0.38] %MEAN % DNSMOS
    
    (0,2) [0.14] %DE
    (1,2) [0.72] %LP
    (2,2) [0.57] %PE
    (3,2) [0.53] %SH  
    (4,2) [0.52] %TM 
    (5,2) [0.00] %UN
    (6,2) [0.44] %MEAN  % (SI-) SDR

    (0,3) [0.79] %DE
    (1,3) [0.65] %LP
    (2,3) [0.79] %PE
    (3,3) [0.80] %SH
    (4,3) [0.55] %TM
    (5,3) [0.77] %UN
    (6,3) [0.74] %MEAN % PESQ

    (0,4) [0.51] %DE
    (1,4) [0.84] %LP
    (2,4) [0.75] %PE
    (3,4) [0.85] %SH
    (4,4) [0.56] %TM
    (5,4) [0.89] %UN
    (6,4) [0.77] %MEAN % SMAQ

    (0,5) [0.73] %DE
    (1,5) [0.89] %LP
    (2,5) [0.86] %PE
    (3,5) [0.61] %SH
    (4,5) [0.62] %TM
    (5,5) [0.78] %UN
    (6,5) [0.77] %MEAN % ViSQOLAudio

    (0,6) [0.71] %DE
    (1,6) [0.95] %LP
    (2,6) [0.90] %PE
    (3,6) [0.80] %SH
    (4,6) [0.76] %TM
    (5,6) [0.95] %UN
    (6,6) [0.87] %MEAN % PEAQ ODG

    (0,7) [0.64] %DE
    (1,7) [0.96] %LP
    (2,7) [0.95] %PE
    (3,7) [0.94] %SH
    (4,7) [0.69] %TM
    (5,7) [0.73] %UN
    (6,7) [0.87] %MEAN % 2f

    (0,8) [0.87] %DE
    (1,8) [0.97] %LP
    (2,8) [0.94] %PE
    (3,8) [0.86] %SH
    (4,8) [0.84] %TM
    (5,8) [0.72] %UN
    (6,8) [0.89] %MEAN % PEAQ-CSM

    (0,9) [0.90] %DE 
    (1,9) [0.90] %LP
    (2,9) [0.93] %PE 
    (3,9) [0.83] %SH 
    (4,9) [0.79] %TM 
    (5,9) [0.94] %UN 
    (6,9) [0.89] %MEAN % NMR

};
\end{axis}
\end{tikzpicture}
}
\caption{Performance in terms of objective/subjective score Pearson's correlation ($R$) \citep{EvalObjective}. The aggregated score AGG summarizes performance across all processing methods.}
\label{tab:heatmap_overall_R}
\end{figure}
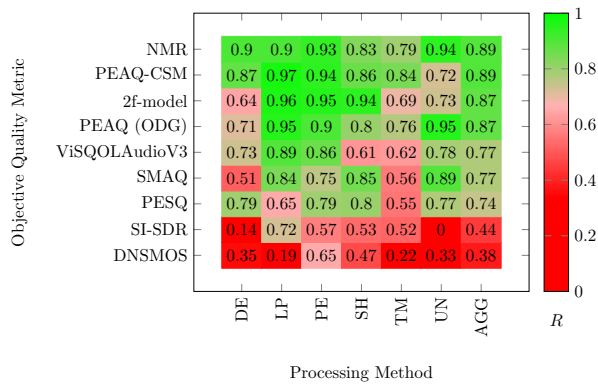

Figure \ref{tab:heatmap_overall_R} summarizes the correlation performances of the considered audio quality metrics.
The six processing methods included in ODAQ allow researchers to diagnose metrics in different quality dimensions and assess their generalization ability.

The top-performing metrics on the current dataset share two key strengths: 1) They are based on auditory perceptual models, and 2) They have been cross-validated across various signal types, including general audio and music. The results support similar findings from previous studies \citep{Torcoli2021, DelgadoPEAQ}. 
Some metrics did not correlate well with the subjective scores in ODAQ, possibly due to the mismatches in application domains or the lack of psycho-acoustic foundations.
This underscores the need for a thorough understanding of a metric's application domain and performance, as well as the importance of cross-validating objective metrics when the application context changes.

Note that the purpose of this benchmark is to present a use case of the expanded dataset instead of evaluating all metrics comprehensively. Since one of the goals of ODAQ is to serve as a resource that is both reproducible and expandable, we expect the dataset to grow and evolve over time. To facilitate such evolution, we provide examples of extending and benchmarking using ODAQ in our online repository \cite{Torcoli2024ODAQ}, and we welcome all potential contributions from the research community.

\section{Summary and Conclusion}

ODAQ provides a valuable resource of processed audio signals with permissive licenses along with subjective scores to address the scarcity of publicly available datasets in the field. We extended ODAQ with subjective scores from newly trained expert listeners and obtained consistent results with previous tests. We also found that listener training can have a substantial impact on the subjective results, especially with respect to how the concept of \emph{anchors} is introduced to the listeners. The results indicate that it may be better to not explicitly bias listeners to the presence of "anchors", and to rather emphasize rating all conditions under test equally with respect to the given scale labels.
Furthermore, we presented a simple use case of ODAQ for benchmarking the performance of objective audio quality metrics.
In conclusion, we believe ODAQ provides a common ground for the research community, and we encourage and welcome further contributions and expansions to this open dataset.

\section{Acknowledgements}
We would like to thank Mhd Modar Halimeh and William Wolcott for their valuable support in creating ODAQ and reviewing this paper. We also warmly thank all participants in the listening tests.

\bibliographystyle{plain}
\begin{small}
\setlength{\bibsep}{0pt plus 0.3ex}
\bibliography{refs}

\begin{thebibliography}{10}

\bibitem{amazon2023}
{About Amazon}.
\newblock Prime video launches a new accessibility feature that makes it easier to hear dialogue in your favorite movies and series, April 2023.
\newblock \url{https://www.aboutamazon.com/news/entertainment/prime-video-dialogue-boost} [Accessed Sep. 2024].

\bibitem{biswas2023}
A.~Biswas and H.~Mundt.
\newblock {AudioVMAF}: Audio quality prediction with {VMAF}.
\newblock In {\em Audio Eng. Soc. (AES) Conv. 155}, Oct. 2023.

\bibitem{brandenburg1987evaluation}
K.~Brandenburg.
\newblock Evaluation of quality for audio encoding at low bit rates.
\newblock In {\em Audio Eng. Soc. (AES) Conv. 82}, March 1987.

\bibitem{Visqol3}
M.~{Chinen}, F.~S.~C. {Lim}, et~al.
\newblock {ViSQOL v3}: An open source production ready objective speech and audio metric.
\newblock In {\em Int. Conf. on Quality of Multimedia Experience (QoMEX)}, 2020.
\newblock code available at \url{https://github.com/google/visqol}.

\bibitem{DelgadoPEAQ}
P.~M. {Delgado} and J.~{Herre}.
\newblock Can we still use {PEAQ}? a performance analysis of the {ITU} standard for the objective assessment of perceived audio quality.
\newblock In {\em Int. Conf. on Quality of Multimedia Experience (QoMEX)}, 2020.

\bibitem{delgado2022data}
P.~M. Delgado and J.~Herre.
\newblock A data-driven cognitive salience model for objective perceptual audio quality assessment.
\newblock In {\em IEEE Int. Conf. Acoustics, Speech and Sig. Proc. (ICASSP)}, pages 986--990, May 2022.

\bibitem{dick2017}
S.~Dick, N.~Schinkel-Bielefeld, and S.~Disch.
\newblock Generation and evaluation of isolated audio coding artifacts.
\newblock In {\em Audio Eng. Soc. (AES) Conv. 143}, Oct. 2017.

\bibitem{ebu_r128}
{EBU~R~128}.
\newblock Loudness normalization and ppermitted maximum level of audio signals.
\newblock European Broadcasting Union ({EBU}), Aug. 2020.

\bibitem{fuchs2012dialogue}
Harald Fuchs, S~Tuff, and C~Bustad.
\newblock Dialogue enhancement - technology and experiments.
\newblock {\em European Broadcasting Union (EBU) Technical Review Q2}, 2012.

\bibitem{herre2023}
J.~Herre and S.~Dick.
\newblock Introducing the free web edition of the "{Perceptual Audio Coders -- What To Listen For"} educational material.
\newblock In {\em Audio Eng. Soc. (AES) Conv. 154}, May 2023.

\bibitem{hu2007evaluation}
Yi~Hu and Philipos~C Loizou.
\newblock Evaluation of objective quality measures for speech enhancement.
\newblock {\em IEEE Trans. Audio, Speech, Language Proc.}, 16(1):229--238, 2007.

\bibitem{MUSHRA}
{ITU-R Rec. BS.1534-3}.
\newblock Method for the subjective assessment of intermediate quality level of audio systems.
\newblock Int. Telecom. Union (ITU), Radiocommunication Sector, Oct. 2015.

\bibitem{PESQ}
{ITU-R Rec. P.862.2}.
\newblock {Wideband Extension to Recommendation P.862 for the Assessment of Wideband Telephone Networks and Speech Codecs}.
\newblock Int. Telecom. Union (ITU), Radiocommunication Sector, December 2007.

\bibitem{PEAQ}
{\relax ITU-R Rec. BS.1387-2}.
\newblock Method for objective measurements of perceived audio quality.
\newblock Int. Telecom. Union (ITU), Radiocommunication Sector, 2023.

\bibitem{EvalObjective}
{\relax ITU-T Rec. P.1401}.
\newblock Methods, metrics and procedures for statistical evaluation, qualification and comparison of objective quality prediction models.
\newblock Int. Telecom. Union (ITU), Radiocommunication Sector, 2012.

\bibitem{n12232}
{ISO/IEC} {JTC1/SC29/WG11}.
\newblock {USAC} verification test report {N12232}.
\newblock Int. Org. Standardisation (ISO), 2011.

\bibitem{Kabal}
P.~Kabal.
\newblock {An Examination and Interpretation of ITU-R BS.1387: Perceptual Evaluation of Audio Quality}.
\newblock Technical report, MMSP Lab Technical Report, McGill University, 2002.
\newblock Code available at \url{http://www-mmsp.ece.mcgill.ca/Documents/Software/}.

\bibitem{ksrWaspaa19}
Thorsten Kastner and Jürgen Herre.
\newblock An efficient model for estimating subjective quality of separated audio source signals.
\newblock In {\em {IEEE Workshop on Applications of Sig. Proc. to Audio and Acoustics (WASPAA)}}, October 2019.
\newblock Implementation details: \url{https://www.audiolabs-erlangen.de/resources/2019-WASPAA-SEBASS}.

\bibitem{TFCTDF}
M.~Kim, W.~Choi, et~al.
\newblock {KUIELab-MDX-Net}: A two-stream neural network for music demixing.
\newblock In {\em Music Demixing Workshop}, Nov. 2021.

\bibitem{roux2018sdrhalfbakeddone}
Jonathan Le~Roux, Scott Wisdom, et~al.
\newblock Sdr -- half-baked or well done?
\newblock In {\em IEEE Int. Conf. Acoustics, Speech and Sig. Proc. (ICASSP)}, pages 626--630, 2019.
\newblock code for mono signals available at: \url{https://github.com/sigsep/bsseval/issues/3}.

\bibitem{Petermann22}
D.~Petermann, G.~Wichern, et~al.
\newblock The cocktail fork problem: Three-stem audio separation for real-world soundtracks.
\newblock In {\em IEEE Int. Conf. Acoustics, Speech and Sig. Proc. (ICASSP)}, pages 526--530, May 2022.

\bibitem{reddy2022dnsmos}
Chandan~KA Reddy, Vishak Gopal, and Ross Cutler.
\newblock Dnsmos p.835: A non-intrusive perceptual objective speech quality metric to evaluate noise suppressors.
\newblock In {\em IEEE Int. Conf. Acoustics, Speech and Sig. Proc. (ICASSP)}, pages 886--890, 2022.
\newblock code available at: \url{https://github.com/microsoft/DNS-Challenge/tree/master/DNSMOS}.

\bibitem{DeepFilterNet2}
H.~Schr\"oter, A.~Escalante-B., et~al.
\newblock {DeepFilterNet2}: Towards real-time speech enhancement on embedded devices for full-band audio.
\newblock In {\em Int. Workshop Acoustic Sig. Enhancement (IWAENC)}, Sep. 2022.

\bibitem{openunmix}
F.~St\"oter, S.~Uhlich, et~al.
\newblock {Open-Unmix - A} reference implementation for music source separation.
\newblock {\em J. Open Source Software}, 4(41):1667, 2019.

\bibitem{Torcoli2021}
M.~Torcoli, T.~Kastner, and J.~Herre.
\newblock Objective measures of perceptual audio quality reviewed: An evaluation of their application domain dependence.
\newblock {\em IEEE/ACM Trans. Audio, Speech, Language Proc.}, 29:1530--1541, 2021.

\bibitem{Torcoli2024ODAQ}
M.~Torcoli, C.-W. Wu, et~al.
\newblock {ODAQ}: Open dataset of audio quality.
\newblock In {\em IEEE Int. Conf. Acoustics, Speech and Sig. Proc. (ICASSP)}, pages 836--340, April 2024.

\bibitem{torcoli2018comparing}
Matteo Torcoli and Sascha Dick.
\newblock Comparing the effect of audio coding artifacts on objective quality measures and on subjective ratings.
\newblock In {\em Audio Eng. Soc. (AES) Conv. 144}, 2018.

\bibitem{torcoli2021dialog}
Matteo Torcoli, Christian Simon, et~al.
\newblock Dialog+ in broadcasting: First field tests using deep-learning-based dialogue enhancement.
\newblock In {\em Int. Broadcasting Conv. (IBC), technical papers}, 2021.

\bibitem{tversky1974judgment}
Amos Tversky and Daniel Kahneman.
\newblock Judgment under uncertainty: Heuristics and biases: Biases in judgments reveal some heuristics of thinking under uncertainty.
\newblock {\em Science}, 185(4157):1124--1131, 1974.

\bibitem{miao_wang_2022_6549559}
Miao Wang, Christoph Boeddeker, et~al.
\newblock Pesq (perceptual evaluation of speech quality) wrapper for python users, 2022.
\newblock code available at \url{https://doi.org/10.5281/zenodo.6549559} and \url{https://github.com/ludlows/PESQ}.

\bibitem{ward2019casualty}
Lauren Ward, Matthew Paradis, et~al.
\newblock Casualty accessible and enhanced (a\&e) audio: Trialling object-based accessible tv audio.
\newblock In {\em Audio Eng. Soc. (AES) Conv. 147}, 2019.

\bibitem{wu2021}
C.~W. Wu, P.~A. Williams, and W.~Wolcott.
\newblock A multitask teacher-student framework for perceptual audio quality assessment.
\newblock In {\em European Sig. Proc. Conf. (EUSIPCO)}, pages 396--400, Aug. 2021.

\bibitem{zielinski:2008}
S.~Zielinski, F.~Rumsey, and S.~Bech.
\newblock On some biases encountered in modern audio quality listening tests-a review.
\newblock {\em J. Audio Eng. Soc. (JAES)}, 56(6):427--451, June 2008.

\end{thebibliography}
\end{small}

\end{document}